\documentclass{aastex63} 

\usepackage[capbesideposition=right]{floatrow}
\usepackage{wrapfig}
\usepackage[right = 0.5in, bottom = 1in]{geometry}

\newcommand{\Rs}{$ R_{\odot} $}
\def\ion[#1 #2]{#1\,{\sc #2}}

\submitjournal{ApJL}
\shortauthors{Boe et al.}

\begin{document}

\title{The Double-Bubble CME of the 2020 December 14 Total Solar Eclipse}

\author{Benjamin Boe}
\affil{ Institute for Astronomy, University of Hawaii, Honolulu, HI 96822, USA}

\author{Bryan Yamashiro}
\affil{ Institute for Astronomy, University of Hawaii, Honolulu, HI 96822, USA}

\author{Miloslav Druckm\"uller}
\affil{Faculty of Mechanical Engineering, Brno University of Technology, Technicka 2, 616 69 Brno, Czech Republic}

\author{Shadia Habbal}
\affil{ Institute for Astronomy, University of Hawaii, Honolulu, HI 96822, USA}

\correspondingauthor{Benjamin Boe}
\email{bboe@hawaii.edu}
\begin{abstract}
Total solar eclipses (TSEs) continue to provide an invaluable platform for exploring the magnetic topology of the solar corona and for studying dynamic events such as Coronal Mass Ejections (CMEs) -- with a higher spatial resolution over a larger spatially continuous extent than is possible to achieve with any other method at present. In this Letter, we present observations of the full extent of a `double-bubble' CME structure from the solar surface out to over 5 solar radii, as captured during the 2020 December 14 TSE. Its evolution through the corona was recorded from two observing sites separated by 13 minutes in their times of totality. The eclipse observations are complemented by a plethora of space-based observations including: Extreme Ultraviolet observations of the solar disk and low corona from {\textit{SDO}}/AIA and {\textit{STEREO-A}}/EUVI, white-light coronagraph observations from  {\textit{SOHO}}/LASCO-C2, radio from  {\textit{STEREO-A}}/WAVES and  {\textit{WIND}}/WAVES, and X-ray from  {\textit{GOES-16}}. We also characterize the magnetic field with a potential field source surface model. This CME event itself is of particular interest, as it demonstrates interactions between a prominence channel and an active region that led to the double-bubble structure. Despite the plethora of space-based observations, only the eclipse data are able to provide the proper context to connect these observations and yield a detailed study of this unique CME. 
\end{abstract} 

\section{Introduction} 
\label{intro}

Despite the short duration of Total Solar Eclipses (TSEs), eclipse observations have often been employed to study Coronal Mass Ejections (CMEs; \citealt{Airapetian1994,Koutchmy2004}). TSE data have been used in conjunction with space-based observations of the solar disk and outer corona before and after the eclipse to investigate CME dynamics \citep{Alzate2017, Druckmuller2017, Boe2018}. Multiple eclipse sites distributed across the path of totality have even been utilized to observe changes in the corona due to an actively propagating CME \citep{Hanaoka2014, Boe2020a}. Nevertheless, the more comprehensive CME studies have used coronagraphs due to their ability to observe the corona semi-continuously for years or even decades (e.g., \citealt{Gopalswamy2001, Robbrecht2009, Bein2011}).

\par
Ground-based coronagraphs cover the lower corona, but they are limited by atmospheric scattering to a maximum of $\sim$1.3 \Rs \ \citep{Boe2021}. On the other hand, space-based coronagraphs are only able to probe the outer region of the corona beyond $\sim$2.2 \Rs\ due to diffraction of photospheric light at the edge of the occulter. Consequently, no ground- or space-based coronagraphic observations can effectively cover the heliocentric distance range of $\sim$1.3 to 2.2 \Rs, except for the LASCO-C1 coronagraph which unfortunately failed in 1998 and produced a limited quantity of data \citep{Zhang2006}. {\textit{STEREO}}/COR1 can also in principle measure this distance range, but the high level of scattered light in the instrument restricts its reliable data to difference images at a very low spatial resolution (e.g., Figs. 1 and 2 in \citealt{Bein2011}). Without access to several \Rs\ of spatially continuous coronal imaging, CME evolution cannot be comprehensively probed, and tracing a CME back to its origin at the Sun remains only marginally reliable. Moreover, \cite{Bein2011} demonstrated that $\sim 74\%$ of CMEs undergo their most dramatic acceleration below 1.5 \Rs, while some exhibit a peak in their acceleration profile at 2 -- 3 \Rs. These findings underscore the importance of accessing the entire distance range from 1--3 \Rs\ for the quantitative evaluation of CME evolution in the corona.

\par

At present, the only platform available to observe CMEs from their sources on the Sun out to an extent of several solar radii simultaneously is with a TSE. Furthermore, recent eclipse observations have demonstrated the exceptionally higher spatial resolution that is achievable with white-light TSE images, in contrast with existing coronagraphs (see \citealt{Alzate2017, Boe2020b, Habbal2021}). The outstanding advantages of TSEs for studies of CME dynamics is further demonstrated in this Letter with white-light observations from the 2020 December 14 TSE  which simultaneously captured the full extent of a double-bubble CME in the corona starting from its tethers at the Sun to over 5 \Rs \ (Section \ref{Eclipse}). Complemented by space-based data from the same day and a potential field source surface (PFSS) model (Section \ref{CME}), the dynamic evolution is shown to be driven by the magnetic connectivity between a prominence, a streamer above it, as well as neighboring active regions on the day of the eclipse.

\section{The 2020 December 14 TSE}
\label{Eclipse}
\begin{figure*}[t!]
\centering
\epsscale{0.7}
\includegraphics[width = 6.5in]{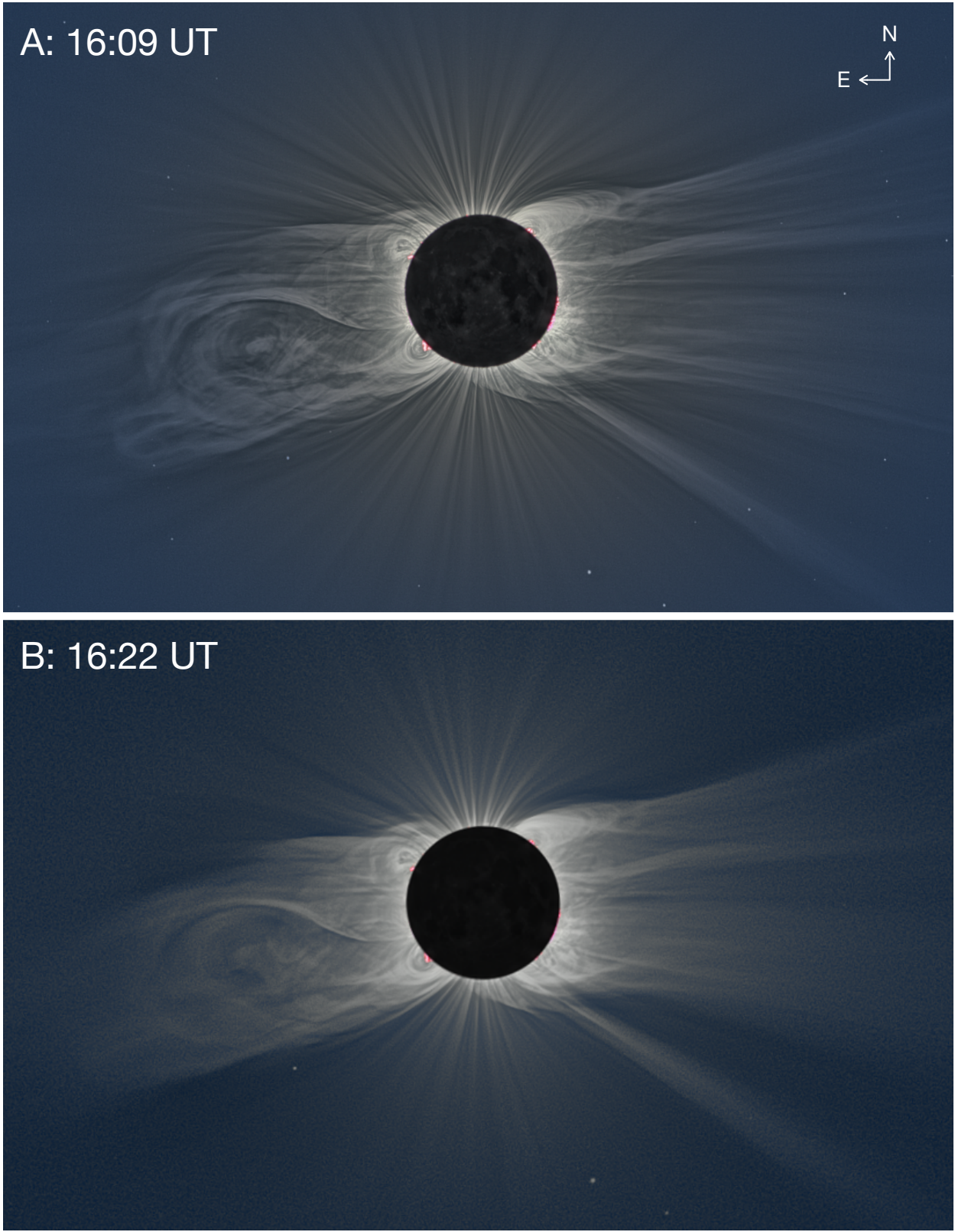}
\caption{A: White-light data from the 2020 December 14 TSE from N\'euquen, Argentina at 16:09 UT (see Section \ref{Eclipse}). Solar cardinal directions are noted in the top right corner. An animation of the CME propagation during the eclipse at N\'euquen from 16:08:37 to 16:09:58 UT is available. The realtime duration of the video is 7 seconds. B: Same as A, but from R\'io Negro, Argentina at 16:22 UT. These images have been processed to enhance fine-scale structures and are not in absolute units.}
\label{figEclipse}
\end{figure*}

White-light images of the corona taken from observers at two different sites along the path of totality during the 2020 December 14 TSE are shown in Figure \ref{figEclipse}. The first to see totality was Andreas M\"oller at Fortin Nogueira, in the province of N\'euquen, Argentina (top panel) and the second was Dario Harari at Bah\'ia Creek, in the province of R\'io Negro, Argentina (bottom panel). The images from N\'euquen were taken with a Nikon Z6, full-frame, mirrorless camera equipped with a 400 mm focal length lens and a 5.6 focal ratio. The resulting image in the top panel of Figure \ref{figEclipse} is a composite of frames with exposure times between 1/640 and 3 seconds. The images from R\'io Negro were taken with a Nikon D7200, APS-C format DSLR camera with a 105 mm focal length lens and a 2.5 focal ratio. The latter image is a composite of frames with exposure times ranging between 1/500 and 1/30 seconds. The composite images were created by aligning and stacking the images using a Fourier transform phase-correlation alignment technique \citep{Druckmuller2009} and processed to enhance structural features \citep{Druckmuller2006}.

\par
\begin{figure*}[t!]
\centering
\epsscale{0.7}
\includegraphics[width = 6.5in]{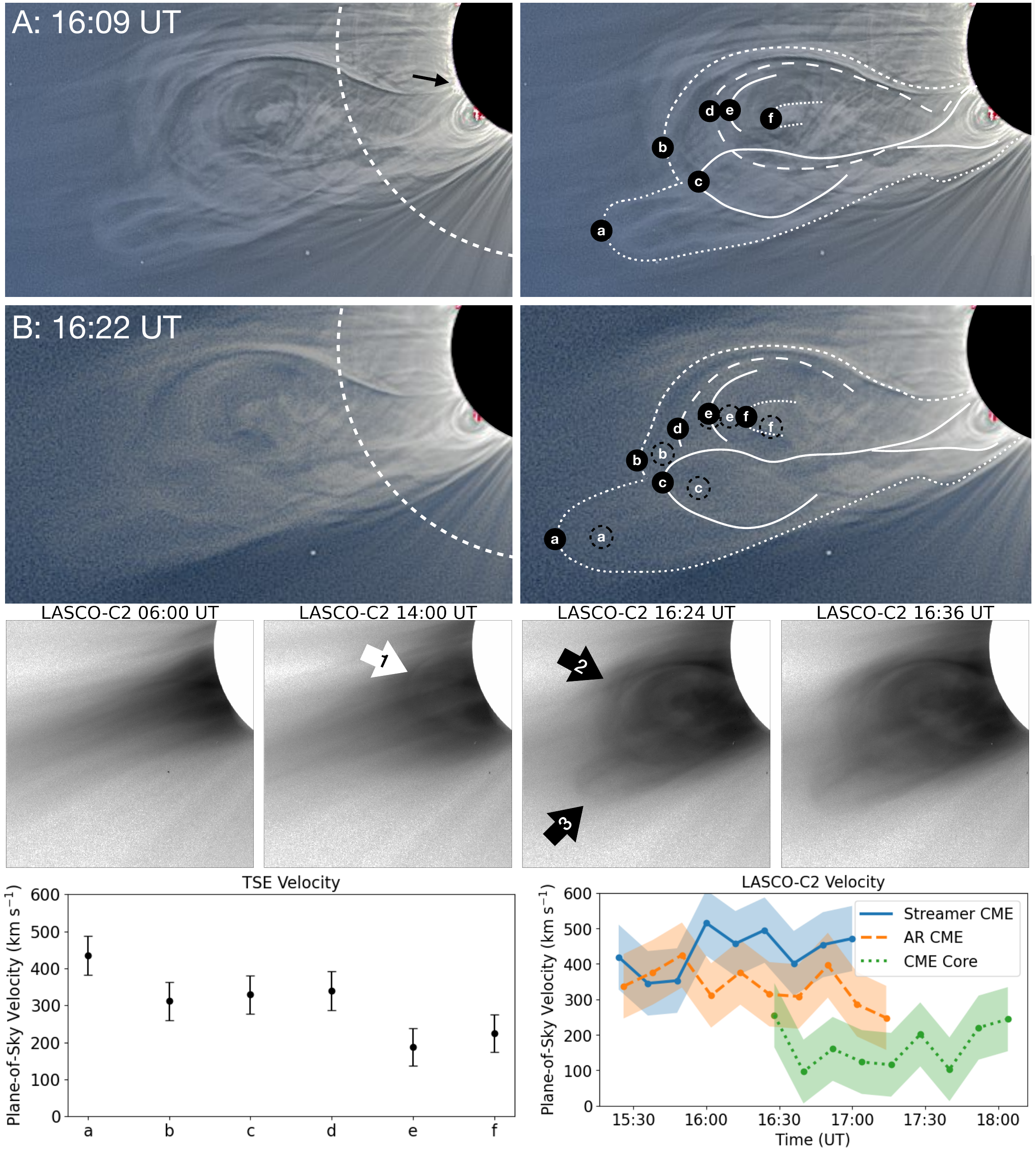}
\caption{The top two rows correspond to the TSE images from N\'euquen (A) and R\'io Negro (B) respectively. The left panels are cropped versions of the images from Figure \ref{figEclipse}, with the dashed arc showing approximately the inner most extent of LASCO-C2 observations (2.2 \Rs). The right panels are the same except with hand-drawn lines indicating the approximate magnetic structure of various features of the CME. Several distinct structures are labeled with black circles in both panels. The initial position of the points at site A are plotted as dashed circles for the site B trace to show their motion between eclipse sites. The middle four panels are color-inverted and cropped images from the LASCO-C2 space-based coronagraph onboard {\textit{SOHO}}. These images have been processed to flatten the radial gradient in brightness to show changing structures in the corona (see Section \ref{LASCO} for processing procedure). The inferred plane-of-sky velocities from the labeled structures in the eclipse panels are shown in the bottom left panel. The velocities inferred for the three main features seen in the LASCO-C2 images are shown in the bottom right panel. An animation showing the full field-of-view of the LASCO-C2 data from 06:00 to 14:54 UT on the day of the eclipse is available online, corresponding to a realtime duration of the video is 19 seconds.}

\label{figCME}
\end{figure*}

The most striking feature in these two eclipse images is a CME which was moving outward from the eastern limb at the time of the eclipse, with a detectable motion during the 13 minutes between the times of totality at each site (shown in the panels of Figure \ref{figEclipse}). Even in the 81 seconds of data from N\'euquen alone, it is possible to see the motion of the CME in contrast to the otherwise static corona. This motion is shown in a video animation which is available in the online version of this Letter, in conjunction with Figure \ref{figEclipse}. To investigate the motion of small scale structures between these two observing sites, we show a zoomed-in version of the images in Figure \ref{figCME}. The CME is clear in both frames, with continuous connectivity seen from the outer shock front all the way back to the footpoint on the solar limb. The CME clearly originated from an active region (AR; indicated by an arrow in the top left panel) which was just north of a large prominence region (see Section \ref{EUV}). The dashed circles in the left panels of Figure \ref{figCME} indicate the approximate inner field-of-view of the LASCO-C2 coronagraph at 2.2 \Rs, demonstrating that LASCO-C2 alone could not directly trace the connectivity of the CME back to its sources on the Sun (see Section \ref{LASCO}). 
\par

The top two right panels of Figure \ref{figCME} contain traces of the main components of the CME activity. The movement of the labeled points over the 13 minutes between sites were used to measure the approximate plane-of-sky (POS) velocities (bottom left panel), which are given as the total magnitude of the velocity vectors on the POS (in 2D). The size of the labels were taken as the 2-$\sigma$ error (i.e., $95\%$ confidence region) for the purposes of determining the uncertainty of the velocities. Given that the active region (and CME associated flare) were located at about E48 longitude, it is reasonable to expect that the CME was moving somewhat toward the Earth, thus the POS velocities represent a lower limit on the true velocity of the CME structures. However, since this CME shows non-radial motions (i.e. deflection from radial), it is not trivial to determine the true 3D propagation direction and speed. 

\par
This CME had a unique topology, with apparently two related yet distinct components, or bubbles. First, there was the classic `ice-cream cone' style CME bubble \citep{Fisher1984} that originated from the active region (see Section \ref{CME}), which had multiple fronts moving at about 300-350~km~s$^{-1}$, while the core behind was moving more slowly at 200~km~s$^{-1}$. Dense cores are routinely observed in CMEs (e.g., \citealt{Gibson1998}), and most likely represent prominence material (i.e. flux rope) from the neutral point of the active region that was at the actual origin of the CME (e.g., \citealt{Manchester2017}). The main part of the CME was clearly deflected northward (about 30$^{\circ}$) from a pure radial expansion (extrapolated from the AR), as the lower sections of the CME were sharply curved upward from the source region on the solar surface. Deflections of CMEs toward the solar equator (i.e. toward the heliospheric current sheet) have been commonly observed, especially near solar minimum (e.g., \citealt{Cremades2006,Zuccarello2012}). Some of the fronts (labels `b' and `d' especially) show a slightly southward propagation, but these fronts appear to have been warped between observations, so it is not clear exactly which region of each front was the same between the frames. The farthest east component of each front is used for the velocity calculation. 
\par

A secondary front and bubble were found to the south of the main CME front. All of the components identified here are clearly part of the same CME eruption, but the unique topology revealed by these eclipse observations indicates that some complex dynamics in the low corona must have led to a variable evolution of different regions of the original CME front. The secondary component of the CME (labels `a' and `c') seemed to be moving at $\sim150$~km~s$^{-1}$ faster (on the POS) than the main CME front, and was at a higher heliocentric distance at the time of the eclipse. This secondary CME component appears to have slightly different footpoints on the Sun compared to the main CME. The eclipse image shows a topological link between these secondary features and the prominence channel that is considerably south of the AR that generates the main CME bubble. 

\section{Spacecraft Data}
\label{CME}

\subsection{LASCO-C2 Coronagraph}
\label{LASCO}
To investigate the propagation of the CME in the corona, we utilize the white-light imaging data from LASCO-C2 on the {\textit{SOHO}} spacecraft \citep{LASCO1995}. The middle panels of Figure \ref{figCME} show a time series of observations made around the time of the eclipse. The LASCO frames presented here are processed to reduce scattered light and cosmic rays. Specifically, we determined the background brightness via the median average of the corona observed 14 days before and after the eclipse. Once the average background is subtracted, a vignetting function is applied (based on the LASCO calibration directory), then cosmic rays are removed using an automatic algorithm which removes transient bright points while leaving background stars. Finally, the intensity is flattened to enhance the outer regions of the corona. A video showing the processed LASCO-C2 data (full field-of-view) is available in the online version of this Letter.
\par

At 6:00 UT on the day of the eclipse, there was a relatively stable equatorial streamer feature that had persisted for several hours. Around 13:30 UT, a slow moving front (as projected on the POS) near the equator on the north end of the streamer (arrow 1 in Fig \ref{figCME}), continued to propagate for a few hours until it was `overtaken' by the fast CME seen during the eclipse, starting at about 15:30 UT. Given its location, speed and timing, this feature is likely to be the same, or part of the same overall structure, as a slowly erupting arch seen in the 30.4 nm data some hours before (see Fig \ref{fig304}, Section \ref{EUV}) and was probably moving behind the POS. By the time of the eclipse (16:09--16:22 UT) and the following couple of hours, the fast CME front moved outward and disturbed a large region of the corona. The panel at 16:24 UT is the most similar to the structure seen during the eclipse, showing the main section of the CME front (arrow 2), as well as the smaller bubble toward the south that was farther from the Sun than the rest of the CME (arrow 3). Even though these structures may appear to have been separate features, they were co-moving through the corona, as seen in both the LASCO-C2 and TSE observations (see Section \ref{Eclipse}). 

\par
Similar to the speed measurements made with the eclipse data (see Fig \ref{figCME}), we measured the locations of some features in the LASCO-C2 images from each frame taken between 15:12 and 18:00 UT. The resolution of LASCO-C2 is considerably worse than the eclipse images, so only three main structures were identified. Namely, the extended front above the prominence streamer, the main CME front that originated from the AR, and the core of the CME. Further, we used a higher uncertainty estimation for the position measurement given the lower resolution. The fastest component of the CME, as seen by LASCO-C2, was the streamer component of the double-bubble eruption which started at $\sim$350~km~s$^{-1}$, then accelerated to $\sim$450-500~km~s$^{-1}$. The AR component of the CME front was moving somewhat slower, at $\sim$300-350~km~s$^{-1}$ and perhaps slowed down slightly as it moved outward. The CME core was moving considerably slower than either of the fronts ($\sim$150-200~km~s$^{-1}$), but showed signs of a slight increase in velocity just before it lost its clearly identifiable structure in the LASCO-C2 data after the eclipse. 

\par
The CME speeds, both from LASCO-C2 and the eclipse data (see Section \ref{Eclipse}) are in agreement with CME catalogs, namely 1) Computer Aided CME Tracking (CACTus; \citealt{Robbrecht2004, Robbrecht2009}) and the 2) Space Weather Database Of Notification, Knowledge, Information (DONKI)\footnote{\url{https://ccmc.gsfc.nasa.gov/donki/}}, which for this CME utilized the NOAA Space Weather Prediction Center CME Analysis Tool (SWPC-CAT; \citealt{Millward2013}). Both catalogs invoke LASCO-C2+C3 observations and sequential frames to derive an estimate of CME parameters such as the speed. CACTus identifies a CME at 15:12 UT with a median velocity of 359~km~s$^{-1}$ and DONKI reports a CME at the same time with a speed of 390~km~s$^{-1}$, both of which are within our uncertainties.

\subsection{Extreme Ultraviolet}
\label{EUV}

To examine the source of the complex CME seen during the eclipse, we use imaging data at EUV wavelengths (30.4 and 17.1 nm) from the AIA instrument onboard the {\textit{Solar Dynamics Observatory (SDO)}} spacecraft \citep{SDO2012} and from the SECCHI-EUVI instrument on the {\textit{Solar Terrestrial Relations Observatory (STEREO-A)} spacecraft \citep{STEREOEUVI2004}. The {\textit{STEREO-A}} spacecraft is in a heliocentric orbit at about 1 AU, and at the time of the eclipse was about 57$^{\circ}$ behind the Earth. {\textit{STEREO-A}} thus provided a more top-down view of the prominence and active region involved in this CME event, similar to the Earth-based view after a solar rotation of about 4.5 days. 

\par
Figure \ref{fig304} shows a time series of observations from the \ion[He ii] 30.4 nm channel of AIA and SECCHI. This channel is useful for investigating properties of the chromosphere and transition region, showing prominences and other relatively low temperature ($\sim 5 \times 10^{4}$ K) coronal features near the Sun. The region where the eclipse CME originated was the southeastern quadrant of the Sun. At 0:00 UT on the day of the eclipse, this region was dominated by two active regions to the north -- ARs 12793 and 12792 which correspond to arrows 1 and 2 respectively in Figure \ref{fig304}. There was also a rather large prominence to the south, including a large hanging cloud of 30.4 nm emission off the limb (arrow 3) somewhat behind the POS (seen from the Earth) indicating that the cloud of material was being suspended and contained by a large closed magnetic field. Finally, there was a main prominence channel closer to the POS (arrows 4 and 5). These features are identified with the numbered arrows in the first {\textit{SDO}}/AIA and {\textit{STEREO-A}}/SECCHI observations in the first two panels of Figure \ref{fig304}. 

\begin{figure*}[t!]
\centering
\epsscale{0.7}
\includegraphics[width =6.5in]{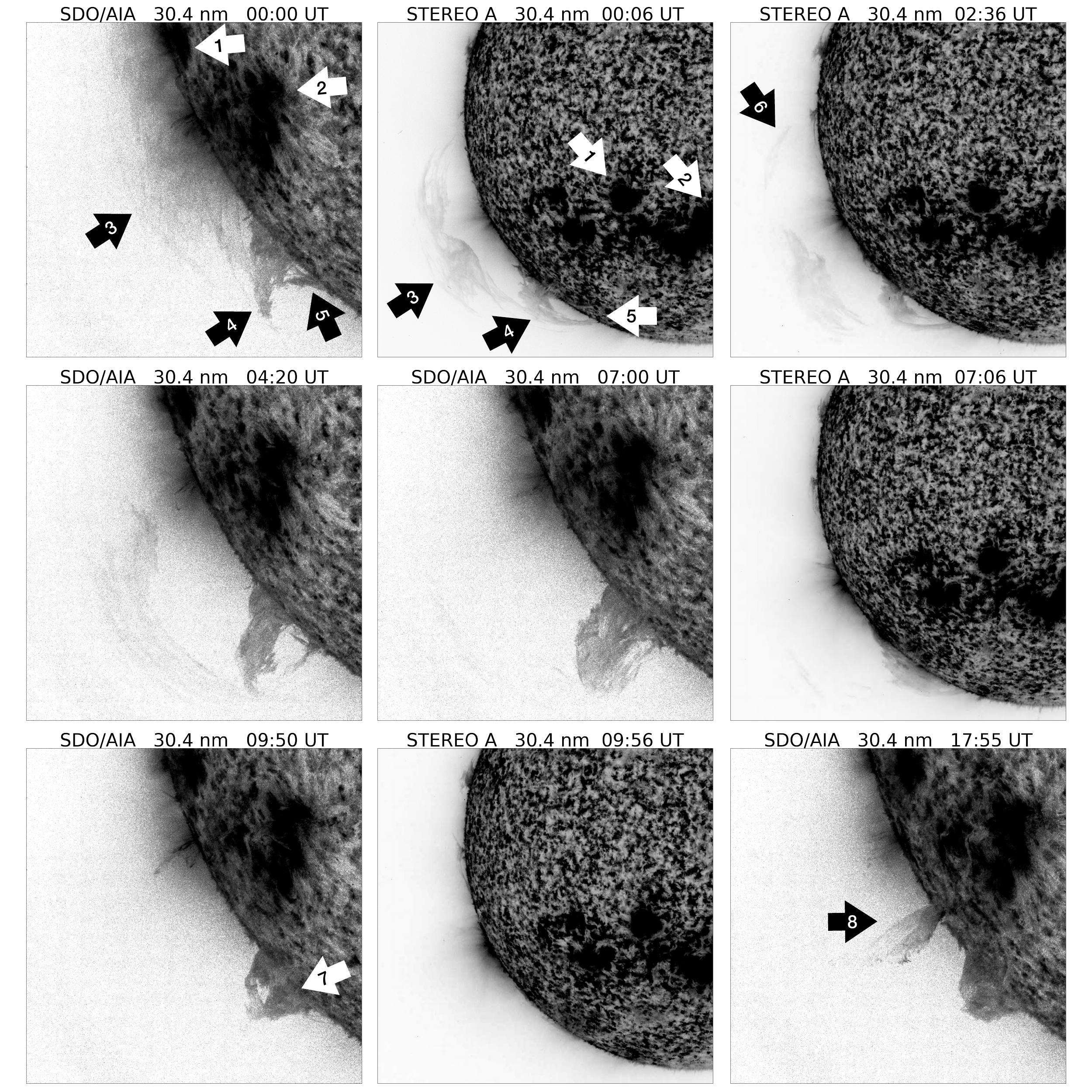}
\caption{Color-inverted and cropped observations of emission from the 30.4 nm (\ion[He ii]) bandpass of {\textit{SDO}}/AIA and {\textit{STEREO-A}}/SECCHI. STEREO-A was 57$^{\circ}$ behind the Earth-view on the day of the eclipse. The data are flattened to enhance the faint emission in the corona relative to the bright emission in the chromosphere. An animation of the the AIA 30.4 nm data from 0:00 to 19:30 UT on the day of the eclipse is available. The realtime duration of the video is 19 seconds. Arrows point to specific structures discussed in Section \ref{EUV}}
\label{fig304}
\end{figure*}

\par

At around 1:00 UT, this arch and cloud structure began moving outward slowly, while the arch showed clear connectivity to the prominence feature close to the POS. As the arch moved outward, it appeared to disconnect from the back side of the Sun (arrow 6) and prominence (arrows 4 and 5) at around 6:00 UT as it disappeared. Such a disappearance is expected for outward moving material seen at EUV wavelengths, since emission lines at these wavelengths are collisionally excited and their emission depends on the square of the electron density (see \citealt{Boe2020a} for more discussion). Shortly after the cloud dissipated, the prominence began to collapse to a smaller size around 8:00 UT, indicating some reconfiguration in the magnetic field that had been suspending the prominence material. By 10:00 UT, the prominence feature behind the POS (seen by SECCHI) had entirely disappeared, while the main prominence (seen by AIA, arrow 7) had reconfigured into a more compact shape. An animation of the AIA 30.4 nm observations from 0:00 to 19:30 UT is available in the online version of this Letter, which shows the evolution of the prominence and AR structures followed by the CME eruption.

\par 
The magnetic connectivity between the AR and prominence on the southeastern limb is illustrated by the \ion[Fe ix] 17.1 nm emission observations from AIA and a corresponding PFSS model, shown in Figure \ref{fig171}. The \ion[Fe ix] emission shows the low regions of the corona at a higher temperature ($\sim 0.5 - 1 \times 10^{6}$ K) than the \ion[He ii] line (i.e. 30.4 nm). We processed the 17.1 nm with the Multi-scale Gaussian Normalization technique (MGN; \citealt{Morgan2014}) to enhance the faint structures in the corona. The PFSS model was generated with pfsspy \citep{Stansby2020}, using the Air Force Data Assimilative Photospheric Flux Transport (ADAPT)\footnote{\url{https://nso.edu/data/nisp-data/adapt-maps/}} synoptic map which uses forward modeling of flux transport and the collection of existing observables to generate a realistic instantaneous synoptic map of the photospheric magnetic field \citep{Hickmann2015}.
\par

\begin{figure*}[t!]
\centering
\epsscale{0.7}
\includegraphics[width = 7in]{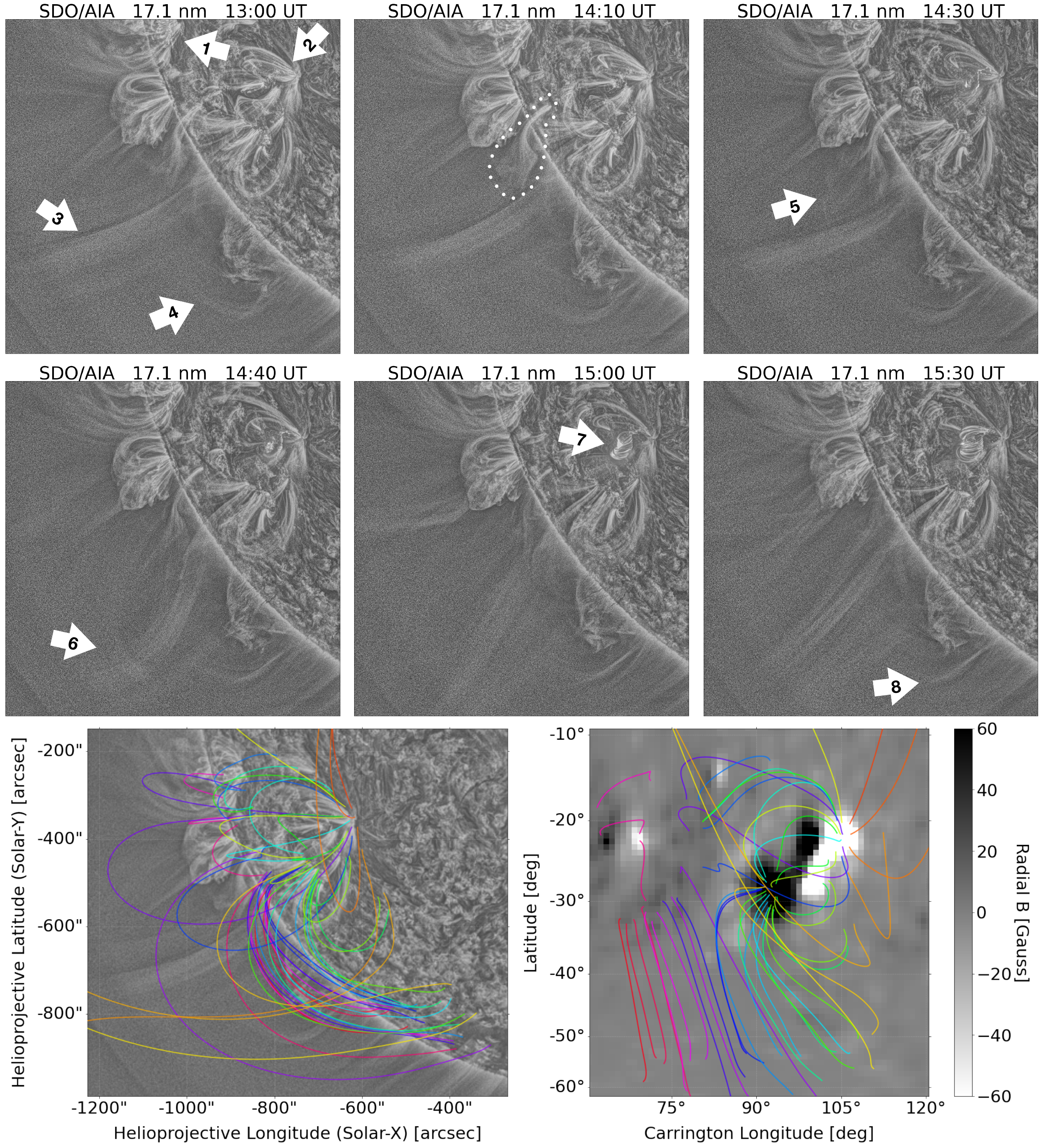}
\caption{MGN processed images from the 17.1 nm (\ion[Fe ix]) bandpass of {\textit{SDO}}/AIA (see Section \ref{EUV}). The field-of-view is identical to that of Figure \ref{fig304}. The bottom panels show traced magnetic field lines from a PFSS model overlaid on the MGN processed 14:00 UT 17.1 nm AIA image (left) and on the ADAPT synoptic magnetic field map (right). Each individual field line trace has a unique color, which is used in both panels. An associated animation of the the AIA data 17.1 nm from 12:00 UT to 16:58 UT on the day of the eclipse is available. The realtime duration of the video is 15 seconds.}
\label{fig171}
\end{figure*}

The 17.1 nm emission was initially rather stable over the morning of the eclipse, showing little change during the 30.4 nm cloud and arch evolution -- except in the immediate vicinity of the large prominence. The ARs were seen clearly in the 17.1 nm emission (arrows 1 and 2 in Fig \ref{fig171}) along with a bright, and apparently open, ray between the closed regions (arrow 3). At the same time, the prominence to the south had a stable hook feature above (arrow 4) which appeared just before 12:00 UT after the collapse and reconfiguration of the prominence as seen in the 30.4 nm data. This morphology persisted for a couple of hours, while the center open ray was moving slightly southward. At 14:30 UT, a structure outside of the main AR arcade is seen to expand (arrow 5 and dotted region) as the CME begins much lower in the AR. This region of the AR arcade specifically appears to be magnetically connected to the prominence channel (see PFSS traces). At 14:40 UT the open ray had bent rapidly toward the prominence (arrow 6), likely interacting with large closed field lines above the prominence. At 15:00 UT there was an emergence and brightening of a cluster of small closed field lines in the center of AR 12792 (arrow 7) confirming the location of the CME origin. This EUV brightening event occurs just after the onset of the radio burst and X-ray flare (see Section \ref{XrayRadio}), indicating magnetic reconnection during the CME. After the CME had passed, the hook feature next to the prominence became increasingly radial, suggesting that the magnetic field around the prominence had reconfigured in the wake of the CME (arrow 8).

\par

The PFSS field line traces in Figure \ref{fig171} reinforce the notion that there was connectivity between the AR and the nearby prominence channel to its south. The traces clearly show that several field lines directly connect the prominence channel to the center of the active region. It is thus reasonable to expect that a CME erupting from the active region could directly perturb the field lines above the prominence. Based on the PFSS model and observed perturbations of both the hook and open ray seen in the 17.1 nm data, it appears that the emerging CME was interacting with the magnetic field immediately above the prominence. During this interaction, it is likely that there was some interchange reconnection between the CME front and the closed-field tethers for the outer layers of the prominence streamer (which connected directly to the AR in the PFSS model), perhaps at a height of about 1.05 to 1.15 \Rs. Specifically the erupting field line outside the AR that is seemingly connected to the prominence channel prior to the CME (identified as the dotted region and arrows 5 and 6 in Fig \ref{fig171}) is likely the main perturbation that leads to the secondary front of the CME seen during the eclipse (see Section \ref{Eclipse}). We explore this idea further in Section \ref{discuss}. 

\subsection{X-ray and Radio}
\label{XrayRadio}

At the same time that the CME is seen erupting from the active region, there was an observed spike in both Radio and X-ray emission. The NOAA Space Weather Prediction Center\footnote{\url{ftp://ftp.swpc.noaa.gov/pub/indices/events/}} reported a C4.0 class X-ray flare and associated optical flare located at about 22$^{\circ}$ south and 48$^{\circ}$ east in heliographic coordinates at 14:09 UT, which is consistent with the location and timing of the erupting AR (see Section \ref{EUV}). The X-ray flux observed on the day of the eclipse by the NASA/NOAA {\textit{Geostationary Operational Environmental Satellite (GOES-16)}} is shown in Figure \ref{figXrayRadio}. We utilize the level 2 EXIS X-Ray Sensor (XRS) 1-minute average data in both short (0.05-0.4 nm) and long (0.1-0.8 nm) wavelength bandpasses. The C4.0 X-ray flare is clearly seen with a peak at 14:37 and ends at 14:56 UT. This peak coincides with the increase of 17.1 nm EUV emission at the center of the erupting active region (see Section \ref{EUV}). 
\par

Concurrent with the X-ray and EUV flares, there was a radio burst observed by {\textit{STEREO}}/WAVES ({\textit{S}}/WAVES \citealt{Bougeret2008}) positioned almost directly above the source of the CME (see Section \ref{EUV}) and at L1 by {\textit{WIND}}/WAVES \citep{Bougeret1995}, as shown in the bottom two panels of Figure \ref{figXrayRadio}. Both instruments capture radio spectra from electric and magnetic field measurements for specified frequencies (Wind/WAVES: 20 kHz to 13.825 MHz, {\textit{S}}/WAVES: 10 kHz to 16.075 MHz) through sets of antennae and receivers. Descriptions of the radio burst data are provided by the NASA Coordinated Data Analysis Web\footnote{\url{https://cdaweb.gsfc.nasa.gov/index.html/}} --  {\textit{WIND}}/WAVES: normalized receiver average voltage; and {\textit{S}}/WAVES: electric field average intensity. 

\par
The radio burst observed at the time of the CME is a type III radio burst that started at high frequency ($>10$ MHz) and propagated down to about 100 kHz for about an hour after the onset of the high frequency burst. Meanwhile, the NOAA Space Weather Prediction center reported a type II burst from ground-based observations (higher frequency than space). The combined type II and type III bursts respectively indicate that the emerging CME created a supermagnetosonic shock in addition to magnetic reconnection in the low corona which liberated high energy electrons from closed-field regions to flow outward on newly opened field lines into the heliosphere (see \citealt{Cane2002}). The nature of this particular radio burst demonstrates that there was magnetic reconnection in the low corona at the time of the CME emergence, which supports the idea that the AR and nearby field lines reconfigured in the immediate wake of the CME event.
\par
Some hours after the eclipse, at about 17:50 UT, a second type III radio burst was seen by the {\textit{S}}/WAVES instrument. This second radio burst corresponds to a second CME which originated from a nearby region. The second CME is seen in the 30.4 nm AIA imaging, as labeled by arrow 8 in Figure \ref{fig304}. The {\textit{WIND}} data did not clearly show this secondary radio burst, and instead displayed some terrestrial auroral kilometric radiation \citep{Gurnett1974}. However, the second type III burst is likely still present in the data, albeit buried by the stronger auroral emission. 

\begin{figure*}[t!]
\centering
\epsscale{0.7}
\includegraphics[width = 6.5in]{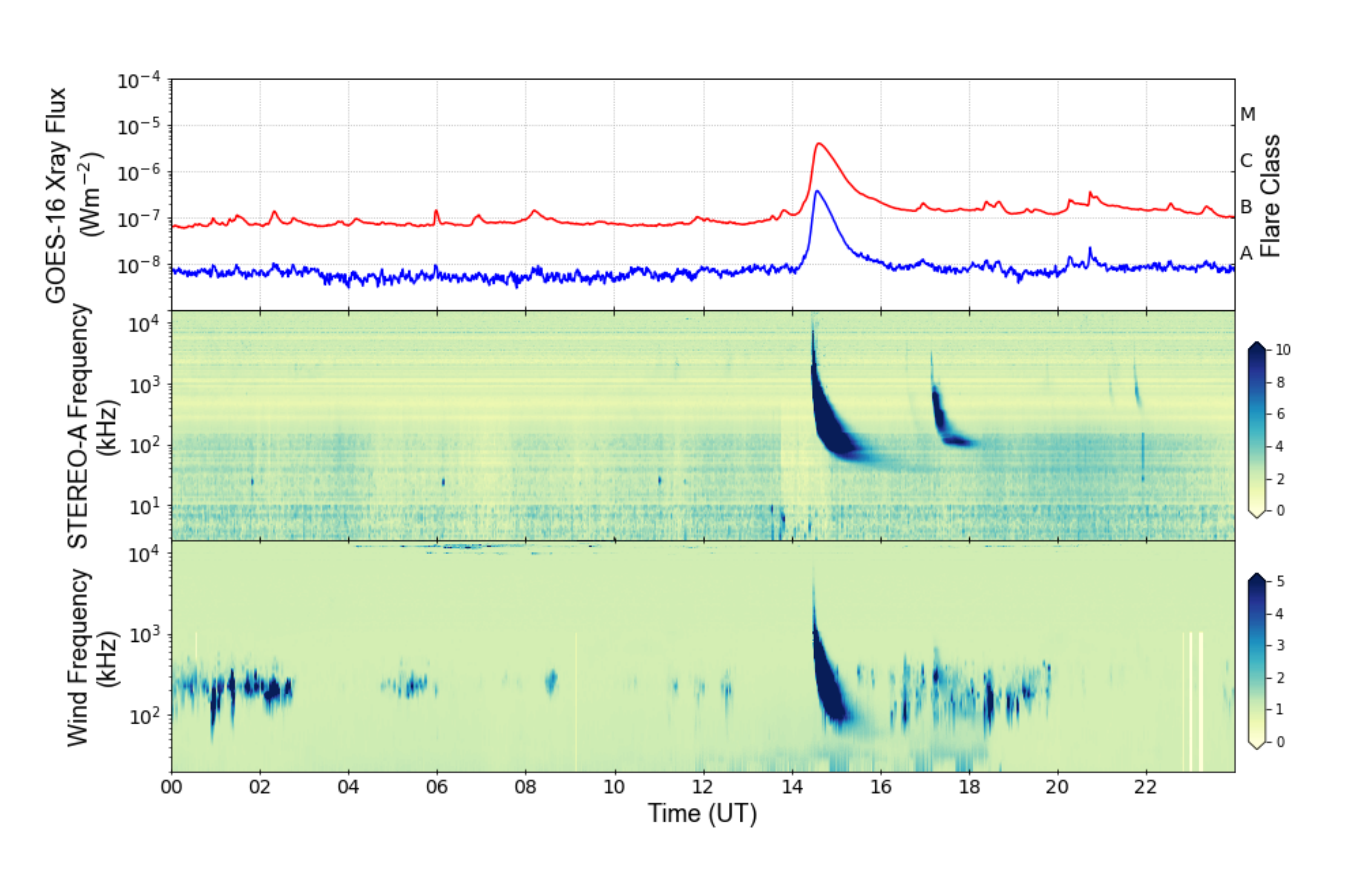}
\caption{Spacecraft X-ray and radio observations from 2020 December 14. Top: C4.0 class X-ray flare detected by {\textit{GOES-16}} in both short (red; 0.05-0.4 nm) and long (blue; 0.1-0.8 nm) wavelength ranges. Middle: {\textit{S}}/WAVES radio observations, showing two type III radio bursts. The stronger burst at about 14:37 UT corresponds to the CME and X-ray flare. Bottom: {\textit{Wind}}/WAVES radio observations from the Earth perspective at L1.}
\label{figXrayRadio}
\end{figure*}

\section{Discussion and Conclusions}
\label{discuss}

In this Letter, we presented white-light data from the 2020 December 14 Total Solar Eclipse (Figure \ref{figEclipse}) which captured the full extent of a CME event spanning several \Rs \ continuously back to its foot points on the Sun. Data were collected at two different sites along the path of totality (see Section \ref{Eclipse}), enabling the inference of changes in the corona over that time. We then investigated the context and evolution of this CME with a combination of space-based emission data of the corona in white-light (see Section \ref{LASCO}), Extreme Ultraviolet (see Section \ref{EUV}), radio and X-ray (see Section \ref{XrayRadio}). We also explored the magnetic connectivity of the region around the CME origin prior to the eruption using a PFSS model (see Fig \ref{fig171}).

\par
The combination of the eclipse data, ancillary observations, and model evidence suggests a narrative of magnetic connectivity between the prominence, the streamer above, and neighboring ARs on the day of the eclipse. The time series of the 30.4 nm emission (see Section \ref{EUV}) indicate that there was a removal and reconfiguration of a large closed-field structure that was connected to the main prominence at around 6:00 UT. This slow eruption was then seen a few hours later as a slow moving front in the LASCO-C2 data (see Section \ref{LASCO}). While this closed-field structure may not have initially interacted directly with the ARs, the reconfiguration of the nearby prominence could have substantially changed the coronal magnetic boundary conditions around the ARs. Indeed, CMEs have often been observed to occur after the ``destabilization of magnetic structures by removing the overlying field" \citep{Lugaz2017}. 

\par

Once this CME emerged into the corona at the time of the eclipse (see Section \ref{Eclipse}), it had a unique structure with two distinct fronts. The top right panels of Figure \ref{figCME} show traces of the magnetic structure, highlighting this double-bubble structure. The high spatial resolution of the eclipse observations (see Fig \ref{figEclipse}) provide important information on the fine-scale topology of the CME, which enables a more detailed inference of its behavior and connectivity compared to the existing spacecraft data. The most likely explanation for the formation of this CME structure is that interchange reconnection (or some similar dynamic) occurred between the prominence streamer and the emerging AR-CME, which had been magnetically connected prior to the CME (based on the PFSS model). The CME then presumably interacted with the prominence channel and pushed some upper layers of prominence streamer (perhaps at about 1.2--1.5~\Rs) outward, which forced them to become a component of the overall CME activity. The notion that there was magnetic reconnection in the low corona that created a double-bubble topology is supported by the simultaneous C4.0 X-ray flare, type II and III radio bursts (see Section \ref{XrayRadio}), and topological changes in the AIA 17.1 nm emission observations (see Fig \ref{fig171}). Moreover, the large prominence region near the disk (Arrow 7 in Fig \ref{fig304}) did not appear to vary much during the CME aside from the slight topology change of the `hook' seen in the 17.1 nm images (arrows 4 and 8 in Fig \ref{fig171}). The lack of any dramatic motion of the prominence indicates that the streamer portion of the double-bubble CME (referred to as front `a' in Section \ref{Eclipse}) was not an independent prominence eruption, but rather was a consequence of reconnection between the streamer above with the main CME bubble (specifically the dotted region and arrows 5 and 6 in Fig \ref{fig171}). 
\par

The magnetic connectivity and reconnection dynamics between neighboring coronal regions, as reported here, are similar to many previous observations of CMEs interacting with neighboring coronal structures. Detailed analysis of recent sympathetic CMEs \citep{Moon2003} have found direct links between the eruption of one CME and another, both in neighboring filament channels \citep{Song2020} and even long range magnetic coupling between active regions \citep{Schrijver2011}. This large-scale connectivity supports the concept that the CME could easily influence a neighboring streamer and filament channel, even if the influence in this case was dominant in the corona rather than near the surface. Coronagraph data and corresponding radio burst signatures have shown that CMEs will collide and magnetically reconnect in the corona \citep{Gopalswamy2001,Lugaz2009,Manchester2017}, and CMEs are commonly deflected in the corona \citep{Cremades2006,Isavnin2014}, including via interactions with streamers \citep{Zuccarello2012}. Further, the topology seen in this eclipse is very similar to the twin-CME cartoon-model proposed by \cite{Li2012}, who were attempting to explain solar energetic particle events originating from a pseudo-streamer CME. While these other CME dynamics are distinct from what is presented here, they all fall into a broader narrative of CMEs interacting with their coronal environment at different stages of their evolution, which is further expanded upon by the findings of this Letter.

\subsection*{Acknowledgments}
We thank Andreas M\"oller and Dario Harari for making their white-light observations at the 2020 Total Solar Eclipse available to us. Financial support was provided to B. Boe by the National Science Foundation under Award No. 2028173. B. Yamashiro was supported by NSF/AURA award No. ND3330. Funding for the DKIST Ambassadors program is provided by the National Solar Observatory, a facility of the National Science Foundation, operated under Cooperative Support Agreement number AST-1400405. M. Druckm\"uller was supported by the Grant Agency of Brno University of Technology, project No. FSI-S-20-6187. This work utilizes data produced collaboratively between AFRL/ADAPT and NSO/NISP.

\bibliographystyle{apj}

\end{document}